\documentclass[10pt]{elsarticle}

\usepackage{amsthm,amsmath,amsfonts,amssymb,amscd,mathrsfs}
\usepackage{txfonts}
\usepackage{supertabular,soul}
\usepackage{tikz, graphicx,color,geometry}
 \usepackage{multirow}
\usepackage[pdftex,
            pdfauthor={Bunimovich},
            pdftitle={Models of Network Growth Based on Specialization of Network Function},
            pdfsubject={Network Theory, Growth Models, Dynamical Systems},
            pdfkeywords={Specialization, Network Growth, Modular and Hierarchical Structure, Intrinsic Stability}]{hyperref}

\usepackage{bbm} 
\usepackage{hyperref}
\usepackage{yfonts}
\usepackage{eucal}
\usepackage{overpic}

\theoremstyle{definition}

\theoremstyle{remark}

\let\oldmarginpar\marginpar
\renewcommand\marginpar[1]{\oldmarginpar[\raggedleft\footnotesize #1]%
{\raggedright\footnotesize #1}}

\begin{document}
\begin{frontmatter}

\title{Uncovering Hierarchical Structure in Social Networks using Isospectral Reductions}


\author[chi]{Chi-Jen Wang}
\address[chi]{Department of Mathematics, National Chung Cheng University, 168 University Road, Min-Hsiung, Chia-Yi 62102, Taiwan, cjwang.math@gmail.com}
\author[seok]{Seokjoo Chae}
\address[seok]{
Department of Mathematical Sciences, Korea Institute of Science and Technology, 291 Daehak-ro, Yuseong-gu, Daejeon 34141, South Korea, sjchea@kaist.ac.kr}
\author[leo]{Leonid A. Bunimovich}
\address[leo]{School of Mathematics, Georgia Institute of Technology, 686 Cherry Street, Atlanta, GA 30332, bunimovich@math.gatech.edu}
\author[ben]{Benjamin Z. Webb}
\address[ben]{Department of Mathematics, Brigham Young University, Provo, UT 84602, USA, bwebb@mathematics.byu.edu}

\begin{abstract}
We employ the recently developed theory of \emph{isospectral network reductions} to analyze multi-mode social networks. This procedure allows us to uncover the hierarchical structure of the networks we consider as well as the hierarchical structure of each mode of the network. Additionally, by performing a dynamical analysis of these networks we are able to analyze the evolution of their structure allowing us to find a number of other network features. We apply both of these approaches to the Southern Women Data Set, one of the most studied social networks and demonstrate that these techniques provide new information, which complements previous findings.
\end{abstract}

\end{frontmatter}



\section{Introduction}\label{sec:1}
The goal of \emph{social network analysis} is to characterize the structure of social relations within a network.  In this analysis it is standard to represent the network as a graph $G=(N,E)$ of nodes $N$ and connecting edges $E$, where the nodes represent network members and an edge represents a connection between two members. The relationship, represented by an edge, depends on the type of network under consideration. For this reason, an edge may also be \emph{weighted} carrying information or data relevant to the network. For instance, in a communication network the frequency or length of communication between pairs of people can be incorporated into the network as the \emph{weights} of the network's edges. In practice, however, network weights are often binary zero-one weights, in which a one represents an \emph{interaction} and a zero \emph{no interaction}.

In a network, the members often consists of different \emph{modes} or distinct types of members. If there is only a single type member the network is referred to as a \emph{single-mode} network. In a two-mode network, often referred to as a \emph{bipartite network}, interactions form between members of two distinct groups. For example, in an actors-to-movies network, each actor is linked to the movies the actor has played in (see, for instance, Watts and Strogatz, 1998). Higher-mode social networks are also possible in which network members are clearly divided into multiple types (cf. Mucha et al.,2010; Murata, 2011).

Beyond these different modes, social networks often exhibit \emph{hierarchical organization}, in which network nodes divide into groups, i.e. \emph{communities}, that further subdivide into smaller groups of groups, and so on over multiple scales. An important aim of social network analysis is to uncover the hierarchical structure of a given network. A typical strategy has been to first break these networks down into smaller groups then to analyze the \emph{position} or relative importance of a member in a given group.

Because of the importance of understanding the hierarchical organization of networks numerous methods have been proposed to determine the position of a member within a given group or, more generally, the network at large (cf. the references in Section \ref{sec:2}). Here, we propose a new approach of uncovering the hierarchical structure of a network. This approach is based on the theory of isospectral network transformations, which allows one to reduce the size of a network while maintaining the network's spectral properties (Bunimovich and Webb, 2014).

The way in which a network is isospectrally reduced depends on a given \emph{criteria} or rule that states which network nodes are \emph{important} and which are less important to the network. Any criteria that selects a specific set of nodes can be used for this purpose, which makes this process quite flexible. A network hierarchy is created by sequentially reducing a given network via this method of isospectral reduction. That is, at each stage in this process the network is reduced. Those nodes that are deemed less important are removed from the network at each stage. Those nodes that remain after each isospectral reduction are considered successively more important than those that removed at a previous stage.

The result is a network hierarchy in which the network members have been sequentially partitioned into a number of groups, each of which is considered more important to the network than the previous group. Naturally, the hierarchy depends on the particular criteria used to select important and unimportant nodes. In this sense the sociologist familiar with a given network is the most natural candidate for developing a useful criteria for obtaining the network's hierarchical structure using this procedure.

In this paper we apply this method of hierarchical analysis to the Southern Women Data Set (Davis, Gardner, and Gardner, 1941, p.148). This data set, yields a two-mode network consisting of eighteen women and the meetings they attended over the course of nine months and is one of the studied networks in social science. We find that the results of our hierarchical analysis of this network are similar in a number of ways to previous results regarding this network but also provide new information that is complementary to these findings.

The  paper is organized as follows. In Section \ref{sec:2} we introduce the Southern Women Date Set and the associated DGG network. In Section \ref{sec:3} we present the results of our analysis describing the hierarchical structure of the DGG network and the single-mode versions of this network based on the method of isospectral reductions. In Section \ref{sec:4} we describe the method used to find the hierarchical structure presented in Section \ref{sec:3}, which include the method of isospectral reduction and the conversion of an $n$-mode network to a smaller single-mode network. In section \ref{sec:5} we describe the procedure of isospectrally reducing a network. In Section \ref{sec:6} we analyze the dynamics the DGG network as it evolves over time. In Section \ref{sec:7} we give some closing remarks.

\section{Southern Women Data Set}\label{sec:2}

One of the earliest social networks to be analyzed is the Southern Women Data Set or the \emph{DGG network} (Davis, Gardner, and Gardner, 1941, p.148). The DGG network is built from fourteen social events attended by eighteen women in 1936 in the town referred to as \emph{Old City}. This data is shown in Table \ref{table1} as an $18\times14$ matrix representing those eighteen women $\mathcal{W}=\{W_1-W_{18}\}$ and which of the fourteen events $\mathcal{E}=\{E_1-E_{14}\}$ they attended. Additionally, the dates of each meeting $\mathcal{E}_1$--$\mathcal{E}_{14}$ were recorded making it possible to analyze the formation and evolution of the network's topology, i.e. analyze the networks \emph{dynamics} (see Section \ref{sec:6}).

\begin{table}[]
\centering
\scalebox{0.85}{
\begin{tabular}{lllllllllllllll}
\hline
\hline
                    & $E_1$  & $E_2$   & $E_3$   & $E_4$   & $E_5$   & $E_6$   & $E_7$   & $E_8$  & $E_9$   & $E_{10}$  & $E_{11}$ & $E_{12}$ & $E_{13}$ & $E_{14}$ \\
\hline
\hspace{0.5cm} Date               & 6/27 & 3/2 & 4/12 & 9/25 & 2/25 & 5/19 & 3/15 & 9/16 & 4/8 & 6/10 & 1/23 & 4/7 & 8/3 & 11/21\\
\hline
$W_1$ (Jefferson) \hspace{1cm}    & 1   & 1    & 1    & 1    & 1    & 1    & 0    & 1   & 1    & 0    & 0   & 0   & 0     & 0   \\
$W_2$ (Mandeville)      & 1   & 1    & 1    & 0    & 1    & 1    & 1    & 1   & 0    & 0    & 0   & 0   & 0     & 0   \\
$W_3$ (Anderson)    & 0   & 1    & 1    & 1    & 1    & 1    & 1    & 1   & 1    & 0    & 0   & 0   & 0     & 0   \\
$W_4$ (B. Rogers)     & 1   & 0    & 1    & 1    & 1    & 1    & 1    & 1   & 0    & 0    & 0   & 0   & 0     & 0   \\
$W_5$ (McDowd)  & 0   & 0    & 1    & 1    & 1    & 0    & 1    & 0   & 0    & 0    & 0   & 0   & 0     & 0   \\
$W_6$ (Anderson)    & 0   & 0    & 1    & 0    & 1    & 1    & 0    & 1   & 0    & 0    & 0   & 0   & 0     & 0   \\
$W_7$ (Nye)    & 0   & 0    & 0    & 0    & 1    & 1    & 1    & 1   & 0    & 0    & 0   & 0   & 0     & 0   \\
$W_8$ (Oglethorpe)      & 0   & 0    & 0    & 0    & 0    & 1    & 0    & 1   & 1    & 0    & 0   & 0   & 0     & 0   \\
$W_9$ (DeSand)       & 0   & 0    & 0    & 0    & 1    & 0    & 1    & 1   & 1    & 0    & 0   & 0   & 0     & 0   \\
$W_{10}$ (Sanderon)     & 0   & 0    & 0    & 0    & 0    & 0    & 1    & 1   & 1    & 0    & 0   & 1   & 0     & 0   \\
$W_{11}$ (Liddell)     & 0   & 0    & 0    & 0    & 0    & 0    & 0    & 1   & 1    & 1    & 0   & 1   & 0     & 0   \\
$W_{12}$ (K. Rogers) & 0   & 0    & 0    & 0    & 0    & 0    & 0    & 1   & 1    & 1    & 0   & 1   & 1     & 1   \\
$W_{13}$ (Avondale)    & 0   & 0    & 0    & 0    & 0    & 0    & 1    & 1   & 1    & 1    & 0   & 1   & 1     & 1   \\
$W_{14}$ (Fayette)      & 0   & 0    & 0    & 0    & 0    & 1    & 1    & 0   & 1    & 1    & 1   & 1   & 1     & 1   \\
$W_{15}$ (Llyod)     & 0   & 0    & 0    & 0    & 0    & 0    & 1    & 1   & 0    & 1    & 1   & 1   & 0     & 0   \\
$W_{16}$ (Murchison)   & 0   & 0    & 0    & 0    & 0    & 0    & 0    & 1   & 1    & 0    & 0   & 0   & 0     & 0   \\
$W_{17}$ (Carleton)    & 0   & 0    & 0    & 0    & 0    & 0    & 0    & 0   & 1    & 0    & 1   & 0   & 0     & 0   \\
$W_{18}$ (Price)     & 0   & 0    & 0    & 0    & 0    & 0    & 0    & 0   & 1    & 0    & 1   & 0   & 0     & 0   \\
\hline
\hline
\end{tabular}}\label{table1}
\vspace{0.1cm}
\caption{The Southern Women Data Set is shown as a zero-one matrix consisting of eighteen rows and fourteen columns, representing the women $\mathcal{W}=\{W_1$--$W_{18}\}$ and events $\mathcal{E}=\{E_1$--$E_{14}\}$ in the data set, respectively. Here, a one in the $ij$--entry of the matrix indicates that the $i$th individual $W_i$ attended the $j$th event $E_j$. A zero indicates an absence at the event. Additionally, the date of each event during the year 1936 is given.}
\end{table}

Because of the relatively small size of this data set and some of the not so obvious patterns it contains, the group (i.e. \emph{community}) and hierarchical structure of this data has been analyzed numerous times. The methods used in this analysis include the use of network correspondence, normalized degree, closeness centrality, betweenness centrality, eigenvector centrality, etc. (Borgatti, 2009; Freeman, 1992; Fielda et al., 2006;  Opsahl, 2013; Doreian et al., 2004; Freeman and White, 1993;  Kuznetsov et al., 2007; Freeman, 2012; Madritscher et al., 2011; Brusco, 2011; Snasel et al., 2009; Xu et al.,  2010; Li and Pang, 2012; Zaversnik et al., 2001; Liu and Murata, 2009; Everett and Borgatii, 2013; Gihosh and Lerman, 2009; Aitkin et al., 2014). Of these findings, twenty-one of them were surveyed by Freeman in 2003 (Freeman, 2003).

In each of these twenty-one investigations an attempt was made to analyze the group structure of the women's social interactions. In eleven of these methods a hierarchical analysis was also given (see Freeman, 2003 for details). The results of this analysis are shown in Table \ref{table:2}, which aside from the last row is a recreation of Figure 10 from Freeman, 2003 p. 25. Each row of Table \ref{table:2} represents a different approach to creating a hierarchy of the women in the Southern Women Data Set. (These are respectively, Davis, Gardner, and Davis, Gardner, and Gardner, 1941 (DGG 41); Homans, 1950 (HOM50); Bonacich, 1978 (BCH78); Doreian, 1979 (DOR79); Bonacich, 1991 (BCH91); Freeman and White, 1993 (FW193); Freeman and White, 1994 (FW293); Borgatti and Everett, 1997 (BE197); Skvoretz, and Faust, 1999 (S\&F99); Roberts, 2000 (ROB00); and Newman, 2001 (NEW01).)

In each row, the women are first divided into two groups then ranked according to a specific analytic procedure. The more important or \emph{core} members are shown to the left in each group. The double vertical lines show the divisions in each method that differentiate core members in each group from \emph{peripheral} members. For instance, the first row of Table \ref{table:2} is the hierarchy devised by the authors of the original study of the Southern Women Data Set. Here the authors specify that women in the first group have the core members $W_1$--$W_4$ followed by the first level of peripheral members $W_5$--$W_7$ then the second level of peripheral members $W_8$ and $W_9$.

In each study represented in Table \ref{table:2} the authors follow this convention of first breaking the women into two (or more possibly overlapping) groups and constructing a hierarchy within each group. In the following section we similarly break the members of the network corresponding to the Southern Women Data Set into groups of core and peripheral members using the method of isospectral reduction. Importantly, this method allows us to rank not only the women in this network but also the events they attended (cf. FW193, FW293, and BE197). The result is a hierarchical analysis of both the women and events in the DGG network.

If we restrict this analysis to just the women in the network and partition them into groups similar to what is done in previous investigations, the result is the hierarchy WCBW17 shown in the last row of Table \ref{table:2}. We comment on the differences and similarities between our hierarchy and the others shown in this table in the following section where the results of our analysis are given.

\begin{table}[]
\centering
\label{table:2}
\begin{tabular}{l | l | l}
\hline
\hline
Analytics &   First Group & Second Group                                                              \\\hline
DGG 41       & 1 2 3 4 $\|$ 5 6 7 $\|$ 8 9                               & 13 14 15 $\|$ 11 12 $\|$ 9 10 16 17 18                        \\
HOM 50       & 1 2 3 4 5 6 7 $\|$ 8                                      & 11 12 13 14 15 $\|$ 8 17 18                                   \\
BCH 78       & 5 $\|$ 1 2 3 4 6                                          & 14 $\|$ 10 11 12 13 15                                        \\
DOR 79       & 1 3 $\|$ 2 4 $\|$ 5 6 7 9                                 & 12 13 14 $\|$ 10 11 15                                        \\
BCH 91       & 5 $\|$ 4 $\|$ 2 $\|$ 1 6 $\|$ 3 $\|$ 7 $\|$ 9 $\|$ 8      & 17 18 $\|$ 12 $\|$ 13 14 $\|$ 1 $\|$ 15 $\|$ 10 $\|$ 16       \\
FW1 93       & 1 2 3 4 $\|$ 5 6 7 8 9 $\|$ 16                            & 13 14 15 $\|$ 10 11 12 17 18 $\|$ 16                          \\
FW2 93       & 1 $\|$ 2 3 4 $\|$ 5 $\|$ 6 $\|$ 7 9                       & 14 $\|$ 12 13 15 $\|$ 11 17 18 $\|$ 10                        \\
BE1 97       & 3 4 $\|$ 2 $\|$ 1 $\|$ 7 $\|$ 6 $\|$ 9 $\|$ 5             & 12 13 $\|$ 11 $\|$ 14 $\|$ 10 $\|$ 15                         \\
S\&F 99      & 1 3 $\|$ 2 $\|$ 4 $\|$ 5 $\|$ 6 $\|$ 7 $\|$ 9 $\|$ 8      & 12 13 $\|$ 14 $\|$ 15 $\|$ 11 $\|$ 10 $\|$ 17 18              \\
ROB 00       & 1 $\|$ 2 $\|$ 4 $\|$ 3 $\|$ 5 $\|$ 6 $\|$ 7 $\|$ 9 $\|$ 8 & 12 $\|$ 13 $\|$ 14 $\|$ 11 $\|$ 15 $\|$ 10 $\|$ 16 $\|$ 17 18 \\
NEW 01       & 1 2 $\|$ 3 $\|$ 4 $\|$ 6 $\|$ 5 $\|$ 7 9                  & 13 14 $\|$ 12 $\|$ 11 $\|$ 15 $\|$ 10 $\|$ 17 18 $\|$ 8 $\|$ 16\\
WCBW 17      & 1 2 3 4 $\|$ 5 6 7 9                                      & 14 $\|$ 13 $\|$ 12 15 $\|$ 10 11 $\|$ 17 18\\
\hline
\hline
\end{tabular}
\caption{The DGG core/periphery membership assignments by the 12 different analytic procedures are shown. The  importance of membership are ranked from left to right and separated  by $\|$ in each procedure. For example, in the case of DGG network (DGG41), the group $\mathcal{G}_1$ is divided into the core members $W_1$--$W_4$, the first peripheral members $W_5$--$W_7$, and the second level of peripheral members $W_8$--$W_9$. The last row shows the hierarchy obtained via the method of isospectral reductions presented in this paper (see Section \ref{sec:3}).}
\end{table}

\section{Results}\label{sec:3}

\begin{figure}
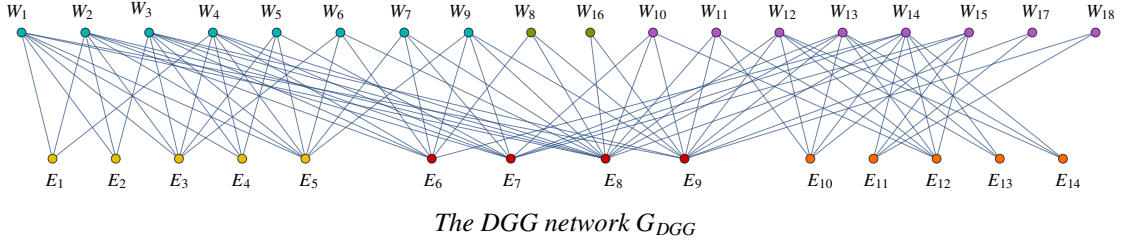

\vspace{0.5cm}
\begin{center}
\begin{overpic}[scale=.4]{DGGNet}
    \put(4.5,0){\scriptsize{$E_1$}}
    \put(10,0){\scriptsize{$E_2$}}
    \put(15.5,0){\scriptsize{$E_3$}}
    \put(21,0){\scriptsize{$E_4$}}
    \put(27,0){\scriptsize{$E_5$}}
    \put(38,0){\scriptsize{$E_6$}}
    \put(45,0){\scriptsize{$E_7$}}
    \put(54,0){\scriptsize{$E_8$}}
    \put(61,0){\scriptsize{$E_9$}}
    \put(72,0){\scriptsize{$E_{10}$}}
    \put(77,0){\scriptsize{$E_{11}$}}
    \put(82.5,0){\scriptsize{$E_{12}$}}
    \put(88,0){\scriptsize{$E_{13}$}}
    \put(94,0){\scriptsize{$E_{14}$}}

    \put(1,15){\scriptsize{$W_1$}}
    \put(6.75,15){\scriptsize{$W_2$}}
    \put(12,15.25){\scriptsize{$W_3$}}
    \put(18,15){\scriptsize{$W_4$}}
    \put(23.5,15){\scriptsize{$W_5$}}
    \put(29,15){\scriptsize{$W_6$}}
    \put(35,15){\scriptsize{$W_7$}}
    \put(40.25,15){\scriptsize{$W_9$}}
    \put(46,15){\scriptsize{$W_{8}$}}
    \put(51.5,15){\scriptsize{$W_{16}$}}
    \put(57,15){\scriptsize{$W_{10}$}}
    \put(62.5,15){\scriptsize{$W_{11}$}}
    \put(68.5,15){\scriptsize{$W_{12}$}}
    \put(74,15){\scriptsize{$W_{13}$}}
    \put(79.5,15){\scriptsize{$W_{14}$}}
    \put(85.5,15){\scriptsize{$W_{15}$}}
    \put(91,15){\scriptsize{$W_{17}$}}
    \put(96.75,15){\scriptsize{$W_{18}$}}

    \put(39,-4){\emph{The DGG network $G_{DGG}$}}
\end{overpic}
\vspace{0.5cm}
\caption{A graphical realization of the DGG network $G_{DGG}$ corresponding to Table \ref{table1} where there is an edge between $W_i$ and $E_j$ if $W_i$ attended $E_j$. Following Freeman and Duquenne (Freeman and Duquenne, 1993), yellow and orange nodes represent the first and second set of group events $\mathcal{E}_1=\{E_1$--$E_5\}$ and $\mathcal{E}_2=\{E_{10}$--$E_{14}\}$, respectively. Red nodes represent the joint meetings $\mathcal{J}=\{E_6$--$E_9\}$. The {blue, purple, and green} nodes represent the first, second, and third groups of women $\mathcal{G}_1=\{W_1$--$W_7,W_9\}$, $\mathcal{G}_2=\{W_{10}$--$W_{15},W_{17},W_{18}\}$, and $\mathcal{G}_3=\{W_8,W_{16}\}$, respectively.}\label{Fig:1}
\end{center}
\end{figure}

The DGG network, shown in Figure \ref{Fig:1} as the graph $G_{DGG}=(\mathcal{W}\cup\mathcal{E},E)$, has thirty two nodes which represent both the eighteen women $\mathcal{W}=\{W_1$--$W_{18}\}$ and fourteen events $\mathcal{E}=\{E_1$--$E_{14}\}$ in the Southern Women Data Set. In the graph there is an edge between $W_i$ and $E_j$ if $W_i$ attend event $E_j$. So that our results can be compared to previous results, these nodes are further divided into six subgroups by color according to the scheme described in (Freeman and Duquenne, 1993). Group one, which consists of the nodes ${\mathcal{G}_1}=\{W_1$--$W_7,W_9\}$ are the women who attended events $\mathcal{E}_1=\{E_1$--$E_5\}$ and events $\mathcal{J}=\{E_6$--$E_9\}$. Events $\mathcal{E}_1$ are referred to as the first set of \emph{group events} and events $\mathcal{J}$ are referred to as the \emph{joint meeting}. Group two, which consists of the nodes $\mathcal{G}_2=\{W_{10}$--$W_{15},W_{17},W_{18}\}$ are the women who attended the second set of group events $\mathcal{E}_2=E_{10}$--$E_{14}$ as well as the joint meetings $\mathcal{J}$. The third group of women $\mathcal{G}_3=\{W_8,W_{16}\}$ are those women who only attended the joint meetings.

In Section 3.1 we first describe the hierarchical structure of the individuals and events in the DGG network obtained using the method of isospectral reductions. In Section 3.2 we go on to describe the hierarchical structure of the two single-mode networks associated with the DGG network.

\subsection{Hierarchical Structures of the DGG Network}
\begin{figure}
\begin{overpic}[scale=.44, angle=90]{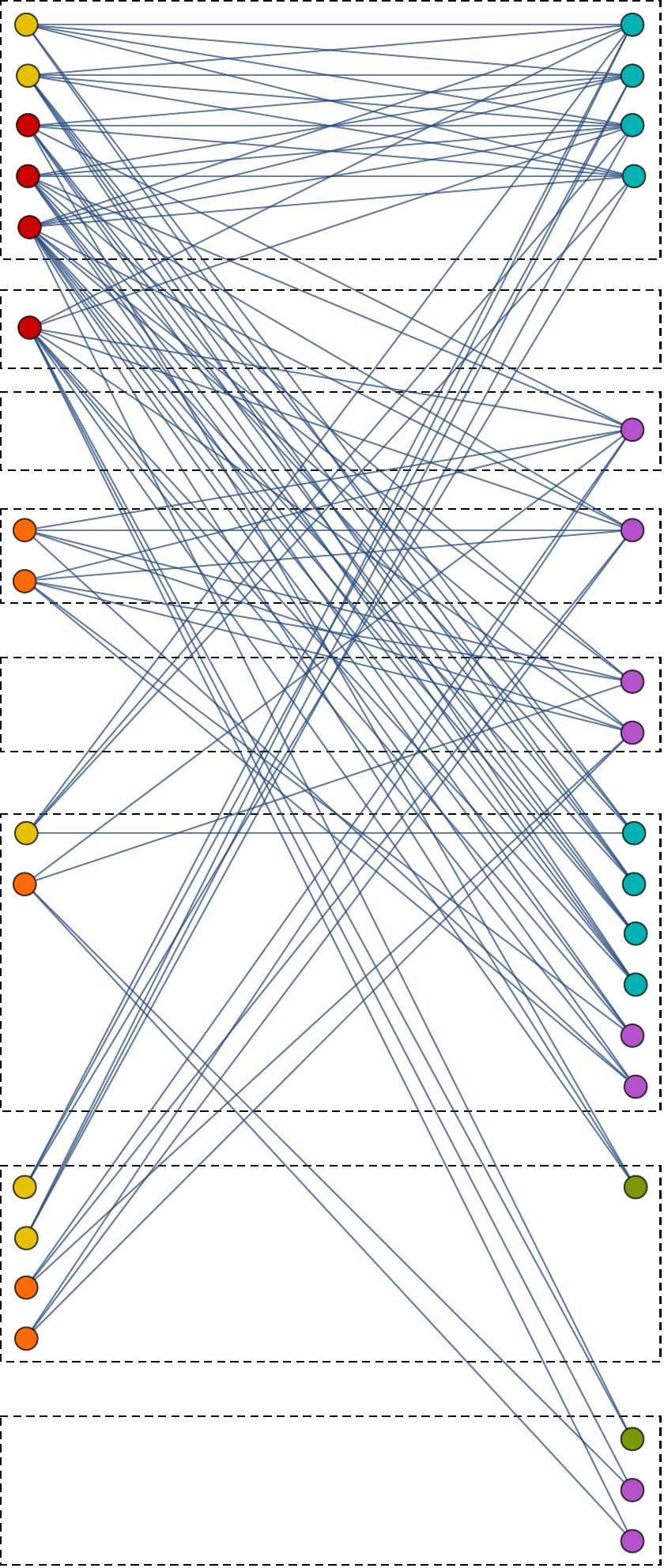}
    \put(1,-2){\scriptsize{$E_3$}}
    \put(4,-2){\scriptsize{$E_5$}}
    \put(7.5,-2){\scriptsize{$E_6$}}
    \put(10.5,-2){\scriptsize{$E_7$}}
    \put(13.5,-2){\scriptsize{$E_8$}}
    \put(20.5,-2){\scriptsize{$E_9$}}
    \put(33,-2){\scriptsize{$E_{10}$}}
    \put(36,-2){\scriptsize{$E_{12}$}}
    \put(52,-2){\scriptsize{$E_{4}$}}
    \put(55,-2){\scriptsize{$E_{11}$}}
    \put(75,-2){\scriptsize{$E_{1}$}}
    \put(78,-2){\scriptsize{$E_{2}$}}
    \put(81,-2){\scriptsize{$E_{13}$}}
    \put(84,-2){\scriptsize{$E_{14}$}}

    \put(1,43){\scriptsize{$W_1$}}
    \put(4,43){\scriptsize{$W_2$}}
    \put(7,43){\scriptsize{$W_3$}}
    \put(10,43){\scriptsize{$W_4$}}
    \put(26.5,43){\scriptsize{$W_{14}$}}
    \put(33,43){\scriptsize{$W_{13}$}}
    \put(42.25,43){\scriptsize{$W_{15}$}}
    \put(45.5,43){\scriptsize{$W_{12}$}}
    \put(52,43){\scriptsize{$W_{5}$}}
    \put(55,43){\scriptsize{$W_{6}$}}
    \put(58.5,43){\scriptsize{$W_{7}$}}
    \put(61.5,43){\scriptsize{$W_{9}$}}
    \put(65,43){\scriptsize{$W_{10}$}}
    \put(68,43){\scriptsize{$W_{11}$}}
    \put(74,43){\scriptsize{$W_{8}$}}
    \put(90.5,43){\scriptsize{$W_{16}$}}
    \put(93.5,43){\scriptsize{$W_{17}$}}
    \put(97,43){\scriptsize{$W_{18}$}}

    \put(6,-6){$h_{core}$}
    \put(21,-6){$h_{1}$}
    \put(27,-6){$h_{2}$}
    \put(35,-6){$h_{3}$}
    \put(44,-6){$h_{4}$}
    \put(61,-6){$h_{5}$}
    \put(80,-6){$h_{6}$}
    \put(94,-6){$h_{7}$}
    \put(30,-10){\emph{Hierarchical Structure of the DGG network}}
\end{overpic}
\vspace{1.2cm}
\caption{{Shown left to right is the hierarchical structure of the DGG network consisting of the eight levels $h_{core}$, $h_{1},\dots,h_7$ of the core and peripheral members of both women $\mathcal{W}=\{W_1-W_{18}\}$ and events $\mathcal{E}=\{E_1-E_{14}\}$.}}\label{Fig:2}
\end{figure}

Here we analyze the hierarchical structure of the DGG network $G_{DGG}$ consisting of both individuals and events.  Our analysis of the DGG network reveals an eight stage hierarchical structure both of the set of women $\mathcal{W}=\{W_1$--$W_{18}\}$ and the set of events $\mathcal{E}=\{E_1$-$E_{14}\}$. Using terminology similar to what has been established in previous investigations of the DGG network we let $h_{core}$, $h_{1},h_{2},\dots$ denote the \emph{core}, and the first, second, etc. \emph{peripheral levels} in the hierarchical structure of the DGG network, respectively. We find that
\begin{align*}
h_{core}&=\{W_1-W_4,E_3,E_5-E_8\};\\
h_{1}&=\{E_{9}\};\\
h_{2}&=\{W_{14}\};\\
h_{3}&=\{W_{13},E_{10},E_{12}\};\\
h_{4}&=\{W_{12},W_{15}\};\\
h_{5}&=\{W_{5}-W_7,W_9-W_{11},E_4,E_{11}\};\\
h_{6}&=\{W_{8},E_1,E_2,E_{13},E_{15}\};\\
h_{7}&=\{W_{16}-W_{18}\};
\end{align*}
form the hierarchical structure of the women and events in the DGG network, obtained using the method of isospectral network reductions (Bunimovich and Webb, 2014). Adopting the notation used in Table \ref{table:2} we write this hierarchy as $H(DGG)=\{h_{core} \ || \ h_1 \ || \ \dots \ || \ h_7\}$, which is shown graphically in Figure \ref{Fig:2}.

This hierarchy differs from previous hierarchies in that it compares both the women $\mathcal{W}$ and events $\mathcal{E}$ of the DGG network (cf. Table \ref{table:2}). However, it can be used to create a hierarchy of the women. By restricting this hierarchy to the women $\mathcal{W}$ we have
\begin{equation}\label{eq:1}
H(DGG|\mathcal{W})=\{1 \ 2 \ 3 \ 4 \ || \ 14 \ || \ 13 \ || \ 12 \ 15 \ || \  5 \ 6 \ 7 \ 9 \ 10 \ 11 \ || \ 8 \ || \ 16 \ 17 \ 18\}
\end{equation}
written as in Table \ref{table:2}. The hierarchy restricted to events is similarly given by
\begin{equation}\label{eq:2}
H(DGG|\mathcal{E})=\{3 \ 5 \ 6 \ 7 \ 8 \ || \ 9 \ || \ 10 \ 12 \ || \ 4 \ 11 \ || \  1 \ 2 \ 13 \ 14\}.
\end{equation}
Although the method of isospectral reduction can be used to break the women of the DGG network into groups, i.e. the core and peripheral members as shown in \eqref{eq:1}, these groups are quite different from those found in previous investigations (cf. Table \ref{table:2}). One reason is because in these earlier studies the women were first placed into groups then the hierarchy of each group was determined. In our method both processes are done simultaneously. However, if we, for instance restrict our hierarchy given in \eqref{eq:1} to the groups ${\mathcal{G}_1}=\{W_1$--$W_7,W_9\}$ and $\mathcal{G}_2=\{W_{10}$--$W_{15},W_{17},W_{18}\}$ indicated in Figure \ref{Fig:1}, the result are the hierarchies
\begin{equation}\label{eq:hier}
H(DGG|\mathcal{G}_1)=\{1 \ 2 \ 3 \ 4 \ \| \ 5 \ 6 \ 7 \ 9\}  \ \ \text{and} \ \ H(DGG|\mathcal{G}_2)=\{14 \ \| \ 13 \ \| \ 12 \ 15 \ \| \ 10 \ 11 \ \| \ 17 \ 18\}
\end{equation}
corresponding to group $\mathcal{G}_1$ and group $\mathcal{G}_2$, respectively. These are the hierarchies shown in the last line of Table \ref{table:2}. Interestingly, for the first group $\mathcal{G}_1$, this hierarchy places the women $W_1-W_4$ ahead of the women $W_5-W_7,W_9$, which is also consistent will nearly all of the hierarchies in Table \ref{table:2} except BCH 78 and BCH 91. Similarly, for the second group $\mathcal{G}_2$, this hierarchy places $W_{12}-W_{14}$ ahead or equal to $W_{11}$ and $W_{15}$, which is consistent with most hierarchies in Table \ref{table:2}.

Another way of analyzing the hierarchies given in Figure \ref{Fig:2} and in \eqref{eq:hier} is to compare them against the more active women and the more popular events of the DGG network. An individual in a network such as the DGG network can be classified as being \emph{more} (\emph{less}) \emph{active} if they attended more (less) events than the average person in the network. Similarly, an event can be classified as \emph{more} (\emph{less}) \emph{popular} if it is attended more (less) than the average event in the network. In the DGG network the more active individuals are $W_1$--$W_4$, $W_{12}$--$W_{15}$ and the more popular events are $E_5$--$E_9$.

One can see that the women and events, shown in Figure \ref{Fig:2}, in each level of the hierarchy are quite consistent with those individuals and meetings that are more active and more popular in the network, respectively. For example, the women that are part of the network core $h_{core}$ are the more active members in the group $\mathcal{G}_1$ having attended an average of 7.5 events each. The women that are part of the peripheral levels $h_1-h_4$ are the more active members of the group $\mathcal{G}_2$ having attended an average of 6.5 events each. The individuals in the lower peripheral levels of the network are increasingly less active having attended an average of 4, 3, and 2 events for those women in $h_5$, $h_6$, and $h_7$, respectively.

A similar trend can be seen to some degree in the network events. Here the events in $h_{core}$, $h_1$, $h_3$, $h_5$, and $h_6$ have an average attendance of 9.2, 12, 5.5, 4, and 3, respectively. One might expect that each of the joint events $\mathcal{J}=\{E_6-E_9\}$ would be part of the core $h_{core}$. However, $E_9$ is not in the core whereas one of the most highly attended meetings of the group $\mathcal{G}_1$, namely $E_3$, is in the core.

It is worth noting that $W_1-W_4$ attended {at least as many more popular events than} less popular events. In contrast, $W_{12}-W_{15}$ each attended fewer of the more popular events than they did the less popular events. Moreover, more than half the participants in the more popular events $E_5-E_8$ are those who are more active. In fact, these more popular events have a higher attendance by the more active individuals than from those that are less active.

\subsection{Hierarchical Structures of the Women-to-Women and Events-to-Events Networks}

\begin{figure}
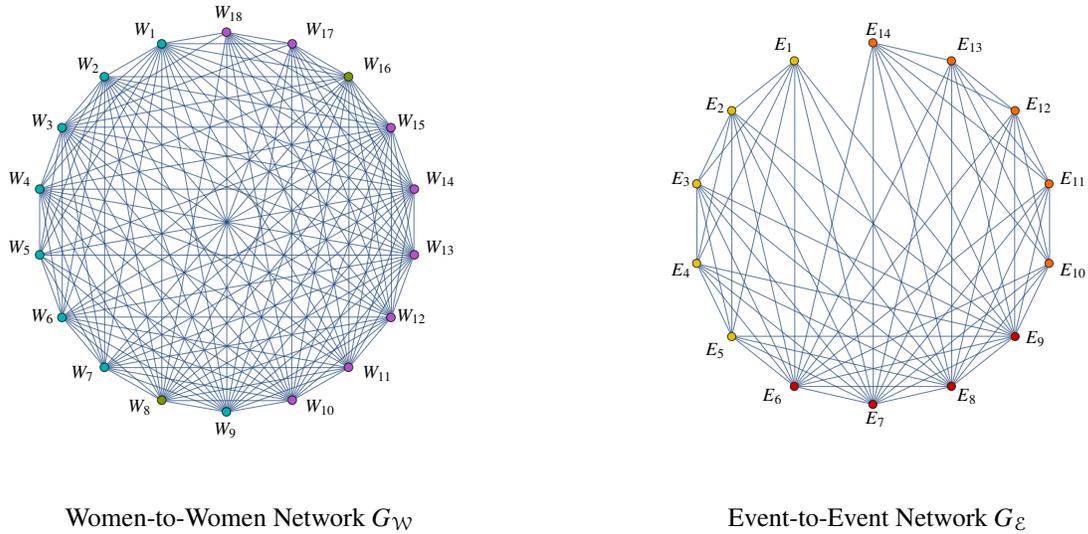

\begin{tabular}{c}
    \begin{overpic}[scale=.40]{SingleModeNets.pdf}
    \put(4,0){Women-to-Women Network $G_\mathcal{W}$}
    \put(62,0){Event-to-Event Network $G_{\mathcal{E}}$}

    \put(66,41.75){\scriptsize{$E_1$}}
    \put(60,36.5){\scriptsize{$E_2$}}
    \put(57,30){\scriptsize{$E_3$}}
    \put(57,22){\scriptsize{$E_4$}}
    \put(60,15){\scriptsize{$E_5$}}
    \put(65,11){\scriptsize{$E_6$}}
    \put(74,9){\scriptsize{$E_{7}$}}
    \put(82,11){\scriptsize{$E_{8}$}}
    \put(88,16){\scriptsize{$E_{9}$}}
    \put(91,22){\scriptsize{$E_{10}$}}
    \put(91,30){\scriptsize{$E_{11}$}}
    \put(88,36.5){\scriptsize{$E_{12}$}}
    \put(82,41.75){\scriptsize{$E_{13}$}}
    \put(74,43){\scriptsize{$E_{14}$}}

    \put(10,43){\scriptsize{$W_1$}}
    \put(5,40){\scriptsize{$W_2$}}
    \put(1,35){\scriptsize{$W_3$}}
    \put(-1,30){\scriptsize{$W_4$}}
    \put(-1,24){\scriptsize{$W_{5}$}}
    \put(1,18){\scriptsize{$W_{6}$}}
    \put(4.5,13){\scriptsize{$W_{7}$}}
    \put(9.5,10){\scriptsize{$W_{8}$}}
    \put(17,8){\scriptsize{$W_{9}$}}
    \put(25,10){\scriptsize{$W_{10}$}}
    \put(30,13){\scriptsize{$W_{11}$}}
    \put(33,18){\scriptsize{$W_{12}$}}
    \put(35.5,24){\scriptsize{$W_{13}$}}
    \put(35.5,30){\scriptsize{$W_{14}$}}
    \put(33,35){\scriptsize{$W_{15}$}}
    \put(30,40){\scriptsize{$W_{16}$}}
    \put(25,43){\scriptsize{$W_{17}$}}
    \put(17,44.5){\scriptsize{$W_{18}$}}
    \end{overpic}
    \end{tabular}
    \vspace{0.1cm}
    \caption{The women-to-women network $G_{\mathcal{W}}$ and event-to-event network $G_{\mathcal{E}}$ associated with the DGG network are shown}\label{Fig:3}
\end{figure}

Aside from finding the hierarchical structure of the DGG network we can also find the hierarchical structure of the single-mode networks that are associated with it. These are the women-to-women and events-to-events networks shown in Figure \ref{Fig:3} as the graphs $G_{\mathcal{W}}=(\mathcal{W},E_{\mathcal{W}})$ and $G_{\mathcal{E}}=(\mathcal{E},E_{\mathcal{E}})$, respectively. The women-to-women network consists of the eighteen women in the DGG network, where there is an edge between two women if they attended the same event at some point. The events-to-events network consists of the fourteen events in the DGG network, where there is an edge between two events if one women attended both events (cf. Figure \ref{Fig:3} and Tables \ref{Tab:3} and \ref{Tab:4}).

The method of isospectral reduction can also be used to find the hierarchical structure of these networks. This results in the hierarchies
\[
H(\mathcal{W})=\{1 \ 2 \ 3 \ 4 \ 6 \ 7 \ 8 \ 9 \ 10 \ 11 \ 12 \ 13 \ 14 \ 15 \ 16 \ || \ 17 \ 18 \ || \ 5\}
\]
for the women-to-women network and
\[
H(\mathcal{E})=\{6 \ 7 \ 8 \ 9 \ || \ 1 \ 2 \ 3 \ 4 \ 5 \ 10 \ 11 \ 12 \ 13 \ 14\}
\]
for the events-to-events network.

The women-to-women network is almost fully connected as can be seen in Figure \ref{Fig:3}. Not surprisingly then, the sequence of {isospectral reductions used to find the core of this network} only removes a few individuals from the network before each of the women has the same number of neighbors and the process terminates. Those that are removed are $W_5,W_{17},W_{18}$. In the event-to-event network the joint meetings $\mathcal{J}=\{E_6-E_9\}$ form the connections between the group meetings $\mathcal{E}_1$ of the first group and the group meetings $\mathcal{E}_2$ of the second group. Because of the role that the joint meetings play in the network these are the events that ultimately form the core of the event-to-event network under isospectral reduction. Together the cores the DGG network and its single-mode reductions are
\[
h_{core}(DGG)=\{W_1\text{--}W_4,E_3,E_5\text{--}E_8\}, \ \ h_{core}(\mathcal{W})=\{W_1\text{--}W_4, W_6\text{--}W_{16}\}, \ \ \text{and} \ \ h_{core}(\mathcal{E})=\{E_6\text{--}E_9\}.
\]

When analyzing a network it is natural to think that the nodes with many neighbors should be part of the network's core. However, in the {full DGG} network the nodes with the {largest} degree are $E_7,E_8$, and $E_9$ but $E_9$ is not part of the network core $h_{core}(F)$. The reason is that the individuals who participated in this event were less active in the overall network. It is worth emphasizing that our analysis of the DGG network is an example illustrating how one can mathematically find the core, or more generally the hierarchical structure of a network, and use this to explain which nodes are most important to the network.

\section{Methods}\label{sec:4}
Suppose a network is given by the graph $G=(N,E)$ in which $N$ is composed of $n$ members. An alternative way of representing the network is by an $n\times n$ square matrix $M$ whose $ij$-entry is the weight of the edge from node $i$ to node $j$ in $G$, and is zero if there is no such edge. The matrix $M=M(G)$ is referred to as the \emph{weighted adjacency matrix} of the network. (If $M$ has zero-one entries then $M$ is referred to as simply the \emph{adjacency matrix} of the network.) The eigenvalues of $M$ are referred to ass the \emph{spectrum} of the network. For example, the adjacency matrix $M(G_{DGG})$ of the DGG network has the block form
\[
M(G_{DGG})=\left[\begin{array}{cc}
0&A\\
A^{T}&0
\end{array}\right]
\]
in which $A$ is the $18\times 14$ matrix of zeros and ones, shown in Table \ref{table1}, describing which individuals attended which events, and where $A^T$ is the \emph{transpose} of the matrix $A$.

The hierarchy describing the structure of the DGG network given in Section \ref{sec:3} is based on the theory of isospectral network transformations (Bunimovich and Webb, 2014), specifically on the theory of isospectral reductions. Using this method one can reduce the size of a network, while preserving the network's spectrum. In this section we first describe this method. We then go on to show that the two-mode DGG network, represented in Table \ref{table1} by a non-square matrix, can be compressed into a single-mode network in two ways, one for each mode of the network. The two resulting networks are the \emph{member-to-member} network and the \emph{event-to-event} network associated with the DGG network, both of which can be isospectrally reduced. We finish the section by describing how any multimode network can be converted to a single-mode network in this manner.


\subsection{Isospectral Network Reductions}

The method of isospectral reduction (Bunimovich and Webb, 2014) can be applied to any network. Via an isospectral network reduction it is possible to uniquely reduce a network given by a graph $G=(N,E)$ over any subset $S$ of its node set $N$. The result is a smaller network given by the reduced graph $\mathcal{R}_S(G)=(S,E_S)$ whose node set is the set $S$ and whose edges $E_S$ are computed using the network's original set of edges $E$. The reason this is referred to as an \emph{isospectral reduction} of the network is that if $M$ is the matrix associated with $G$, the matrix $\mathcal{R}_S(M)$ associated with the reduced network $\mathcal{R}_S(G)$, although smaller, has the same eigenvalues as $M$. (A detailed description of this method is given in Section \ref{sec:5}.)

\begin{figure}
\begin{center}
\begin{tabular}{c}
    \begin{overpic}[scale=.45]{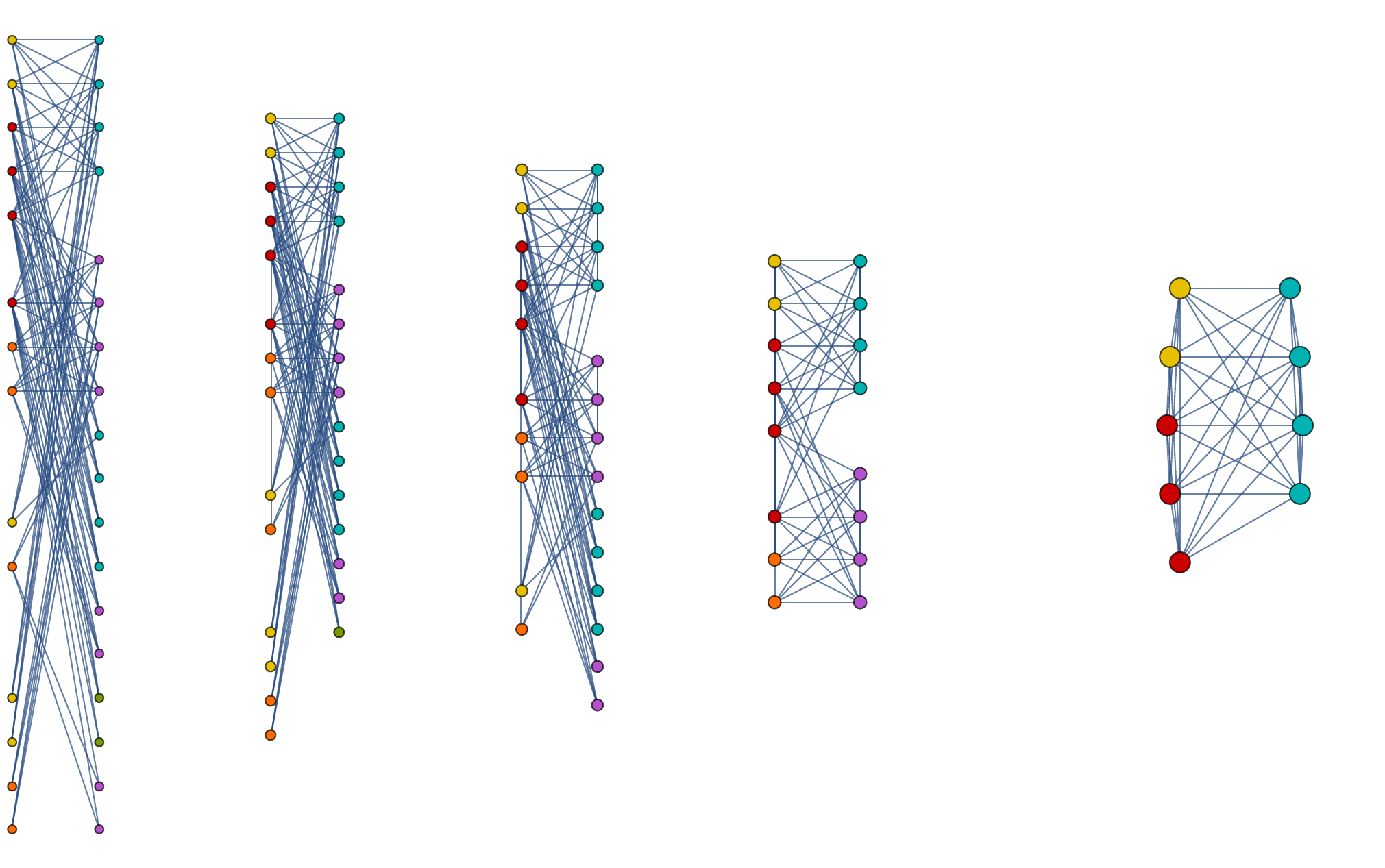}
    \put(69,31.25){\Huge $\dots$}
    \put(0,-1){$R_0=G_{DGG}$}
    \put(22,5){$R_1$}
    \put(40,7.5){$R_2$}
    \put(58,10){$R_3$}
    \put(89,14){$R_7$}

    \put(95,41.5){$W_1$}
    \put(95.5,36.5){$W_2$}
    \put(95.5,31.5){$W_3$}
    \put(95,26.5){$W_4$}

    \put(81,41.5){$E_3$}
    \put(80.5,36.5){$E_5$}
    \put(80.5,31.5){$E_6$}
    \put(81,26.5){$E_7$}
    \put(81,21.5){$E_8$}

    \put(31,-1){(a) Sequence of Isospectral Reductions}
    \put(29,-33.5){(b) Node Degree of the Women in each Reduction}
    \put(29,-61){(c) Node Degree of the Events in each Reduction}
    \end{overpic}\\\\\\\\
    \scalebox{0.85}{\begin{tabular}{lllllllllllllllllll}
    \hline
    \hline
         &  $W_1$  &  $W_2$  & $W_3$ & $W_4$ & $W_5$ &  $W_6$ & $W_7$ & $W_8$ & $W_9$ & $W_{10}$ & $W_{11}$ & $W_{12}$ & $W_{13}$ & $W_{14}$ & $W_{15}$ & $W_{13}$ & $W_{14}$ & $W_{15}$ \\ \hline
$R_0$  & 8 & 7 & 8 & 7 & 4 & 4 & 4 & 3 & 4 & 4 & 4 & 6 & 7 & 8 & 5 & $\textbf{2}_*$& $\textbf{2}_*$ & $\textbf{2}_*$\\
$R_1$  & 8 & 7 & 8 & 7 & 4 & 4 & 4 & $\textbf{3}_*$ & 4 & 4 & 4 & 6 & 7 & 8 & 5 & & & \\
$R_2$  & 10 & 9 & 10 & 9 & $\textbf{4}_*$ & $\textbf{4}_*$ & $\textbf{4}_*$ &  & $\textbf{4}_*$ & $\textbf{4}_*$ & $\textbf{4}_*$ & 5 & 8 & 9 & 5 & & & \\
$R_3$  & 10 & 9 & 10 & 9 &  &  &  &  &  &  &  & $\textbf{7}_*$ & 8 & 9 & $\textbf{7}_*$ & & & \\
$R_4$  & 10 & 9 & 10 & 9 &  &  &  &  &  &  &  &  & $\textbf{7}_*$ & 8 &  & & & \\
$R_5$  & 10 & 9 & 10 & 9 &  &  &  &  &  &  &  &  &  & $\textbf{5}_*$ &  & & & \\
$R_6$  & 10 & 9 & 10 & 9 &  &  &  &  &  &  &  &  &  &  &  & & & \\
$R_7$  & 9 & 9 & 9 & 9 &  &  &  &  &  &  &  &  &  &  &  & & & \\
    \hline
    \hline
\end{tabular}}\\\\\\
\scalebox{0.85}{\begin{tabular}{lllllllllllllll}
    \hline
    \hline
         &  $E_1$  &  $E_2$  & $E_3$ & $E_4$ & $E_5$ &  $E_6$ & $E_7$ & $E_8$ & $E_9$ & $E_{10}$ & $E_{11}$ & $E_{12}$ & $E_{13}$ & $E_{14}$ \\ \hline
$R_0$   & 3 &   3    &  6  &   4    & 8 &   8    & 10 &   14    & 12 &  5     & 4 &  6     & 3 &  3  \\
$R_1$   & $\textbf{3}_*$  & $\textbf{3}_*$ & 6  &  4   & 8  & 8 & 10 &  15 & 12 & 5 & 4  &  6   &  $\textbf{3}_*$ &  $\textbf{3}_*$ \\
$R_2$   &   &   & 6  &  $\textbf{4}_*$ & 10 & 10 & 15 & 12 & 5 & 5 & 6 &  6 &  &      \\
$R_3$   &   &   & 9  &    & 10  &  11     & 14  & 15      & 13  & 8   &   &  9     &   &      \\
$R_4$   &   &   & 9  &    & 10  &  11     & 14  & 14      & 11  & $\textbf{7}_*$   &   &  $\textbf{7}_*$     &   &      \\
$R_5$   &   &   & 9  &    & 10  &  11     & 11  & 11      & 8  &  &   &   &   &      \\
$R_6$   &   &   & 9  &    & 10  &  10     & 10  & 10      & $\textbf{7}_*$ &  &   &   &   &      \\
$R_7$   &   &   & 9  &    & 9  &  9     & 9  & 9      & &  &   &   &   &      \\
    \hline
    \hline
\end{tabular}}\\
    \end{tabular}
    \end{center}
    \caption{(a) The sequence of isospectral reductions $R_1,R_2,\dots,R_7$ of the DGG network $G_{DGG}=R_0$ using the rule that nodes with minimal degree are removed at each step. Edge weights and node labels are neglected, expect for the labels of the core nodes. (b) and (c) The degree of each node in each of the reduced networks is shown. Node(s) with minimal degree are given an asterisk.}\label{Fig:4}
\end{figure}

The main goal of this paper is to demonstrate how isospectral reductions can be used to deduce the hierarchical structure of a network. To demonstrate this we have applied this technique to the DGG network. The specific way in which we have reduce this network is to let $\mu$ be the rule that selects all nodes of a graph that do not have minimal degree, i.e. those nodes that do not have the minimal number of neighbors over the whole graph. For instance, $\mu(G_{DGG})$ are the nodes of the DGG network that have degree more than two, which are all the nodes except $W_{16}-W_{18}$. Using this rule we let $R_0=G_{DGG}$ be our initial unreduced graph and define the $k$th isospectral reduction of $R_0$ to be the graph
\[
R_{k}=\mathcal{R}_{S_{k}}(R_{k-1}) \ \ \text{for} \ \ 1\leq k\leq 7.
\]
This process can be thought of as sequentially removing those nodes that have minimal degree from the graph at each stage this reduction process. It is worth reiterating that this process  does not simply remove those vertices that have minimal degree at every stage if this process, rather an isospectral reduction is used, which preserves the path and cycle structure of the network by modifying the network's weights (see Bunimovich and Webb, 2014 for more details).

The reason this process terminates after seven steps is that $\mu(R_7)$ is empty meaning that every node of this graph has minimal degree. Hence, the graph $R_7$ cannot be further reduced using the rule $\mu$. The results of this sequence of reductions is shown in Figure \ref{Fig:3}(a). The tables in Figure \ref{Fig:3}(b)-(c) show the degree of each node in each reduction $R_0,R_1,\dots, R_7$. Those nodes without an asterisk in $R_{k-1}$ are the nodes that belong to the set $\mu(R_k)$ for $k=1,2,\dots,7$. Those with an asterisk are the nodes removed in the following reduction.

This sequence of reductions describes which of the vertices are most important in the network with respect to the criteria $\mu$, with those that are less important being removed before those that are more important. The way in which we determine which nodes are in the core and peripheral levels of the network is to designate those removed in the first stage of this process as the group $h_7$, those removed in the second stage as the group $h_6$, and so on until those that remain in the final stage make up the core $h_{core}$ of the network.

It is worth emphasizing that the reduction rule $\mu$ used to determine the hierarchy described in Section \ref{sec:3} is only one of many criteria that could be used to find a hierarchy of the DGG network or for that matter any network. In fact, any node characteristic could be chosen as a criteria for creating a network hierarchy including various measures of centrality, number of cycles through a node, etc. Of course, the most natural candidate for developing such a criteria is the sociologist who is familiar with the network. This method is in this sense a flexible and computationally inexpensive method for determining the hierarchical structure of any network of interest.

\begin{table}
\label{Tab:3}
\begin{center}
\scalebox{0.8}{
\begin{tabular}{lllllllllllllllllll}
\hline
\hline
 & $ W_1$ & $W_2$ & $W_3$ & $W_4$ & $W_5$ & $W_6$ & $W_7$ & $W_8$ & $W_9$ & $W_{10}$ & $W_{11}$ & $W_{12}$ & $W_{13}$  & $W_{14}$ & $W_{15}$ & $W_{16}$ & $W_{17}$ & $W_{18}$ \\ \hline
$W_1$ & 8 & 6 & 7 & 6 & 3 & 4 & 3 & 3 & 3 & 2 & 2 & 2 & 2 & 2 & 1 & 2 & 1 & 1 \\
$W_2$ & 6 & 7 & 6 & 6 & 3 & 4 & 4 & 2 & 3 & 2 & 1 & 1 & 2 & 2 & 2 & 1 & 0 & 0 \\
$W_3$ & 7 & 6 & 8 & 6 & 4 & 4 & 4 & 3 & 4 & 3 & 2 & 2 & 3 & 3 & 2 & 2 & 1 & 1 \\
$W_4$ & 6 & 6 & 6 & 7 & 4 & 4 & 4 & 2 & 3 & 2 & 1 & 1 & 2 & 2 & 2 & 1 & 0 & 0 \\
$W_5$ & 3 & 3 & 4 & 4 & 4 & 2 & 2 & 0 & 2 & 1 & 0 & 0 & 1 & 1 & 1 & 0 & 0 & 0 \\
$W_6$ & 4 & 4 & 4 & 4 & 2 & 4 & 3 & 2 & 2 & 1 & 1 & 1 & 1 & 1 & 1 & 1 & 0 & 0 \\
$W_7$ & 3 & 4 & 4 & 4 & 2 & 3 & 4 & 2 & 3 & 2 & 1 & 1 & 2 & 2 & 2 & 1 & 0 & 0 \\
$W_8$ & 3 & 2 & 3 & 2 & 0 & 2 & 2 & 3 & 2 & 2 & 2 & 2 & 2 & 2 & 1 & 2 & 1 & 1 \\
$W_9$ & 3 & 3 & 4 & 3 & 2 & 2 & 3 & 2 & 4 & 3 & 2 & 2 & 3 & 2 & 2 & 2 & 1 & 1 \\
$W_{10}$ & 2 & 2 & 3 & 2 & 1 & 1 & 2 & 2 & 3 & 4 & 3 & 3 & 4 & 3 & 3 & 2 & 1 & 1 \\
$W_{11}$ & 2 & 1 & 2 & 1 & 0 & 1 & 1 & 2 & 2 & 3 & 4 & 4 & 4 & 3 & 3 & 2 & 1 & 1 \\
$W_{12}$ & 2 & 1 & 2 & 1 & 0 & 1 & 1 & 2 & 2 & 3 & 4 & 6 & 6 & 5 & 3 & 2 & 1 & 1 \\
$W_{13}$ & 2 & 2 & 3 & 2 & 1 & 1 & 2 & 2 & 3 & 4 & 4 & 6 & 7 & 6 & 4 & 2 & 1 & 1 \\
$W_{14}$ & 2 & 2 & 3 & 2 & 1 & 1 & 2 & 2 & 2 & 3 & 3 & 5 & 6 & 8 & 4 & 1 & 2 & 2 \\
$W_{15}$ & 1 & 2 & 2 & 2 & 1 & 1 & 2 & 1 & 2 & 3 & 3 & 3 & 4 & 4 & 5 & 1 & 1 & 1 \\
$W_{16}$ & 2 & 1 & 2 & 1 & 0 & 1 & 1 & 2 & 2 & 2 & 2 & 2 & 2 & 1 & 1 & 2 & 1 & 1 \\
$W_{17}$ & 1 & 0 & 1 & 0 & 0 & 0 & 0 & 1 & 1 & 1 & 1 & 1 & 1 & 2 & 1 & 1 & 2 & 2 \\
$W_{18}$ & 1 & 0 & 1 & 0 & 0 & 0 & 0 & 1 & 1 & 1 & 1 & 1 & 1 & 2 & 1 & 1 & 2 & 2\\
\hline
\hline
\end{tabular}
}
\end{center}
\caption{The $18\times 18$ women-to-women matrix $M_{\mathcal{W}}=AA^T$ associated with the DGG network is shown. The $ij$-entry of $M_\mathcal{W}$ is the number of events which both $W_i$ and $W_j$ attended. For $i=j$ the $ij$-entry of $M_{\mathcal{W}}$ represents how many meetings $W_i$ attended.}
\end{table}

\subsection{Conversion to a Single-Mode Network}
Here, we describe how any multimode network can be reduced to a number of smaller single-mode networks. To start, given a non-square $n\times m$ matrix $A$ associated with a two-mode network, the $n\times n$ matrix $A A^T$ is the matrix associated with the first mode of this network whereas the $m\times m$ matrix $A^T A$ is the matrix associated with the second mode of the network. For example, the $18\times 14$ matrix $A$ given in Table \ref{table1} associated with the DGG network can be converted into the $18\times 18$ square matrix $M_{\mathcal{W}}=A A^T$ representing the (weighted) women-to-women network shown in Figure \ref{Fig:3}. The matrix $A$ can also be used to create the $14\times 14$ matrix $M_{\mathcal{E}}=A^T A$ representing the (weighted) event-to-event network also shown in Figure \ref{Fig:3}. These matrices are shown in Tables \ref{Tab:3} and \ref{Tab:4}, respectively. The corresponding networks are shown in Figure \ref{Fig:3}. The hierarchical structure of these networks found in Section 4.1 are computed using the rule $\mu$ also described in Section 4.1.

\[
M=
\left[\begin{array}{ccccc}
0&A_{12}&\dots&A_{13}&A_{1n}\\
A_{21}&0&&&A_{2,n}\\
\vdots&&\ddots&&\vdots\\
A_{n-1,1}&&&0&A_{n-1,n}\\
A_{n,1}&A_{n,2}&\dots&A_{n,n-1}&0

\end{array}\right],
\]
where the transpose $A_{ij}^T=A_{ji}$. For each $i\neq j$ the matrix $M$ \emph{converted} to its $i$-mode over its $j$-mode is the matrix $M^{ij}=A_{ij}A_{ji}$. Hence, there are $n(n-1)/2$ different ways of converting $M$ into the single mode matrix $M^{ij}$, each of which corresponds to a network which can be isospectrally reduced using any rule that is deemed useful.

\begin{table}\label{Tab:4}
\begin{center}
\scalebox{0.8}{
\begin{tabular}{llllllllllllllll}
\hline
\hline
& $ E_1$ & $E_2$ & $E_3$ & $E_4$ & $E_5$ & $E_6$ & $E_7$ & $E_8$ & $E_9$ & $E_{10}$ & $E_{11}$ & $E_{12}$ & $E_{13}$  & $E_{14}$   \\ \hline
$E_1$ & 3 & 2 & 3 & 2 & 3 & 3 & 2  & 3  & 1  & 0 & 0 & 0 & 0 & 0 \\
$E_2$ & 2 & 3 & 3 & 2 & 3 & 3 & 2  & 3  & 2  & 0 & 0 & 0 & 0 & 0 \\
$E_3$ & 3 & 3 & 6 & 4 & 6 & 5 & 4  & 5  & 2  & 0 & 0 & 0 & 0 & 0 \\
$E_4$ & 2 & 2 & 4 & 4 & 4 & 3 & 3  & 3  & 2  & 0 & 0 & 0 & 0 & 0 \\
$E_5$ & 3 & 3 & 6 & 4 & 8 & 6 & 6  & 7  & 3  & 0 & 0 & 0 & 0 & 0 \\
$E_6$ & 3 & 3 & 5 & 3 & 6 & 8 & 5  & 7  & 4  & 1 & 1 & 1 & 1 & 1 \\
$E_7$ & 2 & 2 & 4 & 3 & 6 & 5 & 10 & 8  & 5  & 3 & 2 & 4 & 2 & 2 \\
$E_8$ & 3 & 3 & 5 & 3 & 7 & 7 & 8  & 14 & 9  & 4 & 1 & 5 & 2 & 2 \\
$E_9$ & 1 & 2 & 2 & 2 & 3 & 4 & 5  & 9  & 12 & 4 & 3 & 5 & 3 & 3 \\
$E_{10}$ & 0 & 0 & 0 & 0 & 0 & 1 & 3  & 4  & 4  & 5 & 2 & 5 & 3 & 3 \\
$E_{11}$ & 0 & 0 & 0 & 0 & 0 & 1 & 2  & 1  & 3  & 2 & 4 & 2 & 1 & 1 \\
$E_{12}$ & 0 & 0 & 0 & 0 & 0 & 1 & 4  & 5  & 5  & 5 & 2 & 6 & 3 & 3 \\
$E_{13}$ & 0 & 0 & 0 & 0 & 0 & 1 & 2  & 2  & 3  & 3 & 1 & 3 & 3 & 3 \\
$E_{14}$ & 0 & 0 & 0 & 0 & 0 & 1 & 2  & 2  & 3  & 3 & 1 & 3 & 3 & 3 \\
\hline
\hline
\end{tabular}
}
\end{center}
\caption{The $14\times 14$ event-to-event matrix $M_E=A^TA$ associated with the DGG network is shown. The $ij^{th}$ entry of $M_E$ is the number of women who attended both events $E_i$ and $E_j$. In the case that $i=j$ then the $ij^{th}$ entry of $M_E$ is the entry representing how many women attended $E_i$.}
\end{table}


\section{Isospectral Network Reductions}\label{sec:5}

The tool used throughout this paper to reduce different variations of the full DGG network is the method of isospectral reductions. This theory has been previously applied to the analysis of network stability, survival properties in open dynamical systems, and to gain improved eigenvalue estimates of matrices  to name a few useful applications (Bunimovich and Webb, 2014). In this paper, we have used these techniques to find the {hierarchies and specifically the cores of the DGG and its associated networks, although} these techniques can be applied to any network of interest including the single-mode conversion of multimode networks described in the previous section. In this section we formally describe an isospectral reduction of a network.

Because a graph and its (weighted) adjacency matrix are equivalent, i.e. one can be found from the other, an isospectral network reduction can be described as an isospectral matrix reduction, which is easier to describe than the equivalent notion of an isospectral graph reduction. Suppose that $M$ is the $n\times n$ (weighted) adjacency matrix associated with a network. Given a subset $S$ of the network's nodes, the goal is to reduce this network to a smaller network whose nodes are the set $S$ and whose adjacency matrix, {denoted by $\mathcal{R}_S(M)$}, has the same eigenvalues as the original matrix $M$. This can be done as follows.

For an $n\times n$ matrix $M$ let $N=\{1,\ldots,n\}$. If $R$ and $C$ are subsets of $N$, let $M_{RC}$ be the $|R| \times |C|$ \emph{submatrix} of $M$ with rows indexed by $R$ and columns indexed by $C$. For a subset $S$ of $N$, we let $\bar{S}$ be the compliment of $S$ in $N$. Then the \emph{isospectral reduction} of $M$ over the set $S$ is the $|S|\times|S|$ matrix
\begin{equation}\label{eq:3}
\mathcal{R}_S(M)=M_{SS} - M_{S\bar{S}}(M_{\bar{S}\bar{S}}-\lambda I)^{-1} M_{\bar{S}S},
\end{equation}
where $\lambda$ is a parameter needed to preserve the spectral structure of the matrix under reduction (Bunimovich and Webb, 2014). That is, we reduce the network given by a graph $G$ associated with the matrix $M$ to the smaller network given by the graph $\mathcal{R}_S(G)$ associated with the matrix $\mathcal{R}_S(M)$, which is given by \eqref{eq:3}.

The importance of the set $S$ is that it corresponds to the nodes that remain in the network after the network is reduced. The nodes $\bar{S}$ are those nodes that are removed {from the network as it is reduced}. It is worth noting that the set $S$ can be any subset of the network's nodes so long as $S$ is not empty. In our main example the node set over which we reduced the DGG network were those nodes that did \emph{not} have minimal degree, which we referred to as using the rule $\mu$. This was also the rule used to create hierarchies of the single-mode versions of the DGG network (see Section 3.2).

The way in which the core of a network is found is by sequentially reducing the network over a specific rule. Once a network is reduced with respect to some rule it can again be reduced with respect to the same rule. This sequence of reductions stops once an additional reduction does not modify the graph's structure. This happens precisely when the nodes $S$ chosen by a given rule happen to be the entire node set of the network or the empty set. The importance of this set $S$ is that it forms the \emph{core} of the network with respect to the chosen reduction rule.

The consecutive reductions of a network supply us with additional information since those nodes that are reduced after a number of reductions can be viewed as more important to the network than those that are immediately removed. This allows us to create a hierarchy of nodes based on this sequence of reductions {(cf. Figure 2 and Figure 4)}. Moreover, by combining isospectral reductions with the method described in Section 4.1 it is possible not only to find the hierarchical structure of a network but to find the hierarchical structure of a single mode of an $n$-mode network. Since, there are $n-1$ ways to reduce an $n$-mode network to its $i$th mode, the cores of each of these reductions can be compared with each other with respect to any reduction criteria.


\section{Dynamics}\label{sec:6}

\begin{table}
\centering
\label{table4}
\scalebox{0.85}{
\begin{tabular}{cccccc | ccccc | cccccc}
\hline
\hline
           & 2/25 & 3/2 & 4/12 & 6/27 & 9/25 &                & 3/15 & 4/8  & 5/19 & 9/16 &               &1/23 & 4/7 & 6/10  & 8/3 & 11/21 \\
\hline
               & $E_5$  & $E_2$   & $E_3$   & $E_1$   & $E_4$   &                 & $E_7$   & $E_9$   & $E_6$   & $E_8$  &                 & $E_{11}$ & $E_{12}$   & $E_{10}$ & $E_{13}$ & $E_{14}$ \\ \hline
$W_1$      & 1 & 1 & 1 & 1 & 1 &   & 0 & 1 & 1 & 1 &  &   &   &   &   &   \\
$W_2$      & 1 & 1 & 1 & 1 & 0 &   & 1 & 0 & 1 & 1 &  &   &   &   &   &   \\
$W_3$      & 1 & 1 & 1 & 0 & 1 &   & 1 & 1 & 1 & 1 &  &   &   &   &   &   \\
$W_4$      & 1 & 0 & 1 & 1 & 1 &   & 1 & 0 & 1 & 1 &  &   &   &   &   &   \\
$W_5$      & 1 & 0 & 1 & 0 & 1 &   & 1 & 0 & 0 & 0 &  &   &   &   &   &   \\
$W_6$      & 1 & 0 & 1 & 0 & 0 &   & 0 & 0 & 1 & 1 &  &   &   &   &   &   \\
$W_7$      & 1 & 0 & 0 & 0 & 0 &   & 1 & 0 & 1 & 1 &  &   &   &   &   &   \\
$W_9$      & 1 & 0 & 0 & 0 & 0 &   & 1 & 1 & 0 & 1 &  &   &   &   &   &   \\ \hline
$W_{10}$   &   &   &   &   &   &   & 1 & 1 & 0 & 1 &  & 0 & 1 & 0 & 0 & 0 \\
$W_{11}$   &   &   &   &   &   &   & 0 & 1 & 0 & 1 &  & 0 & 1 & 1 & 0 & 0 \\
$W_{12}$   &   &   &   &   &   &   & 0 & 1 & 0 & 1 &  & 0 & 1 & 1 & 1 & 1 \\
$W_{13}$   &   &   &   &   &   &   & 1 & 1 & 0 & 1 &  & 0 & 1 & 1 & 1 & 1 \\
$W_{14}$   &   &   &   &   &   &   & 1 & 1 & 1 & 0 &  & 1 & 1 & 1 & 1 & 1 \\
$W_{15}$   &   &   &   &   &   &   & 1 & 0 & 0 & 1 &  & 1 & 1 & 1 & 0 & 0 \\
$W_{17}$   &   &   &   &   &   &   & 0 & 1 & 0 & 0 &  & 1 & 0 & 0 & 0 & 0 \\
$W_{18}$   &   &   &   &   &   &   & 0 & 1 & 0 & 0 &  & 1 & 0 & 0 & 0 & 0 \\ \hline
$W_{8}$    &   &   &   &   &   &   & 0 & 1 & 1 & 1 &  &   &   &   &   &   \\
$W_{16}$   &   &   &   &   &   &   & 0 & 1 & 0 & 1 &  &   &   &   &   &   \\
\hline
\hline
\end{tabular}
}
\caption{The reordering of Table 1 {in the horizontal direction first by the event groups $\mathcal{E}_1$, $\mathcal{J}$, and $\mathcal{E}_2$, respectively; then by chronology within these groups. In the vertical direction the table is ordered by the groups $\mathcal{G}_1$, $\mathcal{G}_2$, then $\mathcal{G}_3$. This table give an alternative matrix representation of the DGG network. Blocks that are left blank are blocks of zeros.}}\label{Tab4}
\end{table}

The topology of most real networks are inherently dynamic in that they evolve and change over time. In social networks these structural changes occur as a result of the formation or the dissolving of relationships. The topology of the DGG network is dynamic in that each time there is an event the network grows. Specifically, at each new event new nodes are added to the network, which represent the event and any individuals that attended an event for the first time. The edges added to the network after each event indicate which individuals attended this event.

Berger-Wolf and Saia have previously studied the dynamics of this network using Metagroups (Berger-Wolf and Saia, 2006). In their study the authors investigated the network's dynamics by {chronologically ordering the network's events and analyzing how the similarity between two events $E_i$ and $E_j$ evolves over time}, where the similarity is measured by the generalized \emph{Jaccard measure} (see Berger-Wolf and Saia, 2006 for details).

\begin{figure}
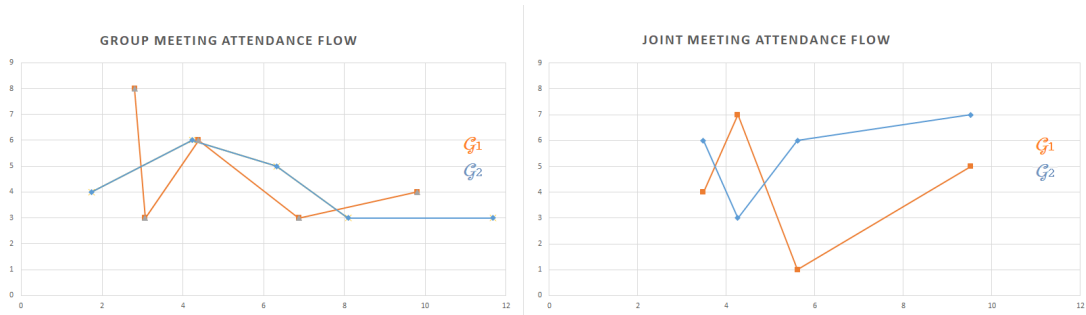

\begin{center}
\begin{tabular}{c}
    \begin{overpic}[scale=.675]{DGGflow}
    \end{overpic}
    \end{tabular}
    \end{center}
    \caption{The total attendance of the two groups $\mathcal{G}_1$ and $\mathcal{G}_2$, shown in red and blue, respectively, during 1936. The horizontal-axis indicates the months in 1936 and the vertical axis the number of attendees. Left: The attendance of $\mathcal{G}_1$ and $\mathcal{G}_2$ to their respective group meetings $\mathcal{E}_1$ and $\mathcal{E}_2$. Right: The attendance of $\mathcal{G}_1$ and $\mathcal{G}_2$ to the joint meetings.}\label{Fig:last}
\end{figure}

Similarly, we begin our analysis by first restricting our attention to the set of events $\mathcal{E}_1$, $\mathcal{E}_2$, and then to the joint events $\mathcal{J}$. We then order the events within these sets chronologically as shown in Table \ref{Tab4}. As can be seen in the table, groups $\mathcal{G}_1$ and $\mathcal{G}_2$, who participated in events $\mathcal{E}_1$ and $\mathcal{E}_2$ respectively, demonstrate behaviors that are quite different. In particular, the women in the first group $\mathcal{G}_1$ all attended their first group meeting $E_5$ but thereafter the group's total attendance to the group meetings forms an oscillating pattern given by the sequence $8$, $3$, $6$, $3$, $4$. In contrast, the second group $\mathcal{G}_2$ shows {an essentially monotone decline in total attendance after it peaks at their second meeting at six attendees. The number of total attendance for this second group at each event is given by $4$, $6$, $5$, $3$, $3$ (see Figure \ref{Fig:last}).}

Notably, at the joint meetings members of the groups $\mathcal{G}_1$ and $\mathcal{G}_2$ also demonstrate {distinct behaviors. The total attendance of $\mathcal{G}_1$ at each of the joint meetings, given by the sequence $6$, $3$, $6$, $7$ has an average of 5.5 and a variance of 3. In contrast, the total attendance of $\mathcal{G}_2$ at each of the joint meetings, given by the sequence $4$, $7$, $1$, $5$ had a lower average of 4.25 and a higher variance of 6.25 indicating that the individuals in $\mathcal{G}_2$ attended the group meetings less frequently and were more sporadic in their attendance.}

The two individuals in the third group $\mathcal{G}_3=\{W_8,W_{18}\}$ are distinguished from the other members of the DGG network in that they only attended the joint meetings. In fact, speaking chronologically, both women missed the first event and attended the second and fourth events of the joint meetings. The only difference in their attendance is that $W_{8}$ attended the third meeting while $W_{16}$ did not.

\section{Discussion}\label{sec:7}

In this paper we have analyzed both the dynamics and structure of the DGG network. To analyze the structure of this network we have employed the method of isospectral network reductions to determine the hierarchical structure of the DGG network as well as the structure of the smaller single-mode women-to-women and events-to-events networks associated with this network. This has allowed us to find the DGG network's core members and events, in various ways (see Section \ref{sec:3}).

Beyond determining the core members of the DGG network, this method {of sequentially reducing a given network} also allows us a way of ranking the importance of any network node depending on the particular criteria we use. In this paper the particular criteria we use to create a hierarchy of the DGG and associated network is the rule that removes nodes with a minimal number of neighbors, via the method of isospectral reductions. Importantly, this method is fundamentally different from simply removing these nodes from the network.  It also adds edges (connections) between remaining nodes, which preserves information regarding the nodes that were removed (Bunimovich and Webb, 2014). By sequentially removing nodes, we rank their importance in that those that are removed earlier are less important than those that are removed later.

Significantly, the method is quite flexible in that any criteria for selecting a unique set of nodes from a network can be used to rank the importance of the network's nodes. From the point of view of applications, such a rule can be used to compare the hierarchical structure of a number of related networks, e.g. a number of different social networks. In fact, various rules can be used to generate and compare different hierarchies of the same network. What is important though is that such rules be designed by experts in the particular field, e.g. sociologists, so that the resulting hierarchies are meaningful.

\section{Acknowledgements}
This work is partially supported by NSF grant DMS-1600568 by the DoD grant HDTRA1-15- 0049.

\section{References}
\hangafter 1
\hangindent 1.5em
\noindent
Aitkin, M., Vu, D., Francis, B., 2014. Statistical modeling of the group structure of social networks.
Social Networks 38, 74-87.

\hangafter 1
\hangindent 1.5em
\noindent
Berger-Wolf, T., Saia, J., 2006. A framework for analysis of dynamic social networks. Paper
presented at the 12th ACM SIGKDD International conference on Knowledge discovery and data
mining (KDD'06), August 20–23, 2006, Philadelphia, Pennsylvania, USA.

\hangafter 1
\hangindent 1.5em
\noindent
Bonacich, P., 1978. Using boolean algebra to analyze overlapping memberships. Sociological Methodology 101-115.

\hangafter 1
\hangindent 1.5em
\noindent
Bonacich, P., Domhoff, G.W., 1981. Latent classes and group membership. Social Networks 3, 175-196.

\hangafter 1
\hangindent 1.5em
\noindent
Bonacich, P., 1991. Simultaneous group and individual centralities. Social Networks 13, 155–168.

\hangafter 1
\hangindent 1.5em
\noindent
Borgatti, S., 2009. Two-mode concepts in social network analysis. Encyclopedia of Complexity
and Systems Science, Springer, 8279-8291.

\hangafter 1
\hangindent 1.5em
\noindent
Borgatti, S., Everett, M., 1997. Network analysis of 2-mode data. Social Networks 19, 243-269.

\hangafter 1
\hangindent 1.5em
\noindent
Borgatti, S. P. and Everett, M. G., 1997. Network analysis of 2-mode data. Social Networks 19:243-269.

\hangafter 1
\hangindent 1.5em
\noindent
Borgatti, S., Halgin,D., 2011. Analyzing affiliation networks. The Sage Handbook of Social Network Analysis, SAGE Publications, Chapter 28.

\hangafter 1
\hangindent 1.5em
\noindent
Brusco, M., 2011. Analysis of two-mode network data using nonnegative matrix factorization.
Social Networks 33, 201-210.

\hangafter 1
\hangindent 1.5em
\noindent
Bunimovich, L., Webb, B., 2012.  Isospectral graph transformations, spectral equivalence, and global stability of dynamical networks.
Nonlinearity 25, 211–254.

\hangafter 1
\hangindent 1.5em
\noindent
Bunimovich, L., Webb, B., 2014. Isospectral Transformations: A New Approach to Analyzing Multidimensional Systems and Networks, Springer.

\hangafter 1
\hangindent 1.5em
\noindent
Cerinsek, M., Batagelj, V., 2015. Generalized two-mode cores. Social Networks 42, 80-87.

\hangafter 1
\hangindent 1.5em
\noindent
Davis, A., Gardner, B., Gardner, M.R., 1941. Deep South; a social anthropological study of caste
and class, University of Chicago Press.

\hangafter 1
\hangindent 1.5em
\noindent
Doreian, P., 1979. On the delineation of small group structure. In H. C. Hudson, ed. Classifying Social Data, San Francisco: Jossey-Bass.

\hangafter 1
\hangindent 1.5em
\noindent
Doreian, P., Batagelj, V., Ferligoj, A., 2004. Generalized blockmodeling of two-mode network data.
Social Networks 26, 29-53.

\hangafter 1
\hangindent 1.5em
\noindent
Everett, M.G., Borgatti, S.P., 2013. The dual-projection approach for two-mode networks. Social
Networks 35, 204-210.

\hangafter 1
\hangindent 1.5em
\noindent
Faust, K., 1997. Centrality in affiliation networks. Social Networks 19, 157-191.

\hangafter 1
\hangindent 1.5em
\noindent
Fielda, S., Frankb, K., Schillerc, K., Riegle-Crumba, C., Mullera, C., 2006. Identifying positions
from affiliation networks: Preserving the duality of people and events. Social Networks 28, 97-123.

\hangafter 1
\hangindent 1.5em
\noindent
Freeman, L., 1992. The sociological concept of "Group": An empirical test of two models. The
American Journal of Sociology 98, 152-166.

\hangafter 1
\hangindent 1.5em
\noindent
Freeman, L., 2003. Finding social groups: A meta-analysis of the southern women data.
Dynamic Social Network Modeling and Analysis: Workshop Summary and Papers.

\hangafter 1
\hangindent 1.5em
\noindent
Freeman, L., 2012. Social network visualization, Methods of. Computational Complexity,
Springer, 2981-2998.

\hangafter 1
\hangindent 1.5em
\noindent
Freeman, L., Duquenne, V., 1993. A note on regular colorings of two mode data. Social Networks
15, 437-441.

\hangafter 1
\hangindent 1.5em
\noindent
Freeman, L., White, D., 1993. Using Galois lattices to represent network data. A note on regular colorings of two mode data.
Sociological Methodology  23, 127-146.

\hangafter 1
\hangindent 1.5em
\noindent
Freeman, L. C. and D. R. White 1994 Using Galois lattices to represent network data. In Sociological Methodology 1993. P. Marsden, ed. Pp. 127-146. Cambridge, MA:
Blackwell.

\hangafter 1
\hangindent 1.5em
\noindent
Gihosh, R., Lerman, K., 2009. Structure of heterogeneous networks. Paper presented at the 2009
International Conference on Computational Science and Engineering (CSE 2009), August 29-31,
2009, Vancouver, Canada.

\hangafter 1
\hangindent 1.5em
\noindent
Homans, G. C., 1950. The Human Group. New York: Harcourt, Brace and Company.

\hangafter 1
\hangindent 1.5em
\noindent
Isah, H., Neagu, D., Trundle, P., 2015. Bipartite network model for inferring hidden ties in crime data. Paper presented at the 2015 IEEE/ACM International Conference on Advances in Social Networks Analysis and Mining (ASONAM 2015), August 25-28, 2015, Paris, France.

\hangafter 1
\hangindent 1.5em
\noindent
Kuznetsov, S., Obiedkov, S., Roth, C., 2007. Reducing the representation complexity of
lattice-based taxonomies.  ICCS 2007, LNAI 4604, 241-254.

\hangafter 1
\hangindent 1.5em
\noindent
Latapy, M., Magnien, C., Vecchio, N., 2008. Basic notions for the analysis of large two-mode networks. Social Networks 30, 31-48.

\hangafter 1
\hangindent 1.5em
\noindent
Li, K., Pang, Y., 2012. A vertex similarity probability model for finding network community structure.
 Advances in Knowledge Discovery and Data Mining. PAKDD 2012. Lecture Notes in Computer Science, vol 7301. Springer, Berlin, Heidelberg.

\hangafter 1
\hangindent 1.5em
\noindent
Liu, X., Murata, T., 2009. Community detection in large-scale bipartite networks. Paper presented at the 2009
IEEE/WIC/ACM International Conference on Web Intelligence,  September  15-18, 2009, Milan, Italy.

\hangafter 1
\hangindent 1.5em
\noindent
Moedritscher, F., Taferner, W., Soylu, A., Causmaecker, P., 2011. Visualization of networked collaboration in digital ecosystems through two-mode network patterns.
Paper presented at the 2011 International ACM Conference on Management of Emergent Digital EcoSystems (MEDES 2011) , November 21-24, 2011, San Francisco, California, USA.

\hangafter 1
\hangindent 1.5em
\noindent
Mucha, P. J., Richardson, T., Macon, K., Porter, M. A., $\&$ Onnela, J. P. (2010). Community structure in timedependent, multiscale, and multiplex networks. Science,
328(5980), 876-878.

\hangafter 1
\hangindent 1.5em
\noindent
Murata, T. (2011). A New Tripartite Modularity for Detecting Communities. Computer Software, 28(1), 154-161.

\hangafter 1
\hangindent 1.5em
\noindent
Newman, M., 2001. The structure of scientific collaboration networks. Proceedings of the National Academy of Science 98:404-409.

\hangafter 1
\hangindent 1.5em
\noindent
Newman, M., 2003. The structure and function of complex networks. SIAM Review 45, 167-256.

\hangafter 1
\hangindent 1.5em
\noindent
Opsahl, T., 2013. Triadic closure in two-mode networks: Redefining the global and local clustering
coefficients. Social Networks 35, 159-167.

\hangafter 1
\hangindent 1.5em
\noindent
Opsahl, T., Panzarasa, P., 2009. Clustering in weighted networks. Social Networks 31, 155-163.

\hangafter 1
\hangindent 1.5em
\noindent
Roberts, J. M. Jr., 2000. Correspondence analysis of two-mode networks. Social Networks 22:65-72.

\hangafter 1
\hangindent 1.5em
\noindent
Skvoretz, J. and Faust, K., 1999. Logit models for affiliation networks. Sociological Methodology 29:253-280.

\hangafter 1
\hangindent 1.5em
\noindent
Snasel, V., Horak, Z., Kocibova, J., Abraham, A., 2009. Reducing social network dimensions using matrix factorization methods.  Paper presented at the  2009 International Conference on
Advances in Social Network Analysis and Mining (ASONAM 2009), July 20-22, 2009, Athens, Greece.

\hangafter 1
\hangindent 1.5em
\noindent
Wasserman, S., Faust, K., 1994. Social network analysis: Methods and applications. Cambridge University Press, New York.

\hangafter 1
\hangindent 1.5em
\noindent
Watts, D., Strogatz, S., 1998. Collective dynamics of 'small-world' networks. Nature 393, 440-442.

\hangafter 1
\hangindent 1.5em
\noindent
Xu, K., Tang, C., Li, C., Jiang, Y., Tang, R., 2010. An MDL approach to efficiently discover
communities in bipartite network. DASFAA 2010, Part I, LNCS 5981, 595–611.

\hangafter 1
\hangindent 1.5em
\noindent
Zaversnik, M., Batagelj, V., Mrvar, A., 2011. Analysis and visualization of 2-mode networks. Paper
presented at the 6th Austrian, Hungarian, Italian and Slovenian Meeting of Young Statisticians,
October 5-7, 2001, Ossiach, Austria.

\end{document}